\documentclass[%
 aip,
 amsmath,
 amssymb,
 reprint,%
]{revtex4-1}

\usepackage{graphicx}% Include figure files

\usepackage{float}
\usepackage{subfigure}
\usepackage[colorlinks,breaklinks,linkcolor=blue,anchorcolor=blue,citecolor=blue,urlcolor=blue]{hyperref}
\usepackage{dcolumn}% Align table columns on decimal point
\usepackage{bm}% bold math
\usepackage[utf8]{inputenc}
\usepackage[T1]{fontenc}
\usepackage{mathptmx}
\usepackage{etoolbox}
\usepackage{tikz}
\usetikzlibrary{shapes}

\makeatother
\begin{document}

\preprint{AIP/123-QED}

\title[]{Dissipative stabilization of high-dimensional GHZ states for neutral atoms}
% Force line breaks with \\
\author{Yue Zhao}
\homepage{These authors contributed equally to this work.}
 \affiliation{ 
Center for Quantum Sciences and School of Physics, Northeast Normal University, Changchun 130024, China}%
\author{Yu-Qing Yang}%
\homepage{These authors contributed equally to this work.}
\affiliation{ 
Center for Quantum Sciences and School of Physics, Northeast Normal University, Changchun 130024, China%\\This line break forced with \textbackslash\textbackslash
}%

\author{Weibin Li}
\email[]{Authors to whom correspondence should be addressed: weibin.li@nottingham.ac.uk and shaoxq644@nenu.edu.cn}
\affiliation{School of Physics and Astronomy, and Centre for the Mathematics and Theoretical Physics of Quantum Non-equilibrium Systems, The University of Nottingham, Nottingham NG7 2RD, United Kingdom%\\This line break forced% with \\
}%
\author{Xiao-Qiang Shao}
\email[]{Authors to whom correspondence should be addressed: weibin.li@nottingham.ac.uk and shaoxq644@nenu.edu.cn}
\affiliation{ 
Center for Quantum Sciences and School of Physics, Northeast Normal University, Changchun 130024, China%\\This line break forced with \textbackslash\textbackslash
}%
\affiliation{Center for Advanced Optoelectronic Functional Materials Research, and Key Laboratory for UV Light-Emitting Materials and Technology of Ministry of Education, Northeast Normal University, Changchun 130024, China}
\date{\today}% It is always \today, today,

\begin{abstract}
High-dimensional quantum entanglement characterizes the entanglement of quantum systems within a larger Hilbert space, introducing more intricate and complex correlations among the entangled particles' states. The high-dimensional Greenberger-Horne-Zeilinger (GHZ) state, symbolic of this type of entanglement, is of significant importance in various quantum information processing applications. This study proposes integrating a neutral atom platform with quantum reservoir engineering to generate a high-dimensional GHZ state deterministically. Leveraging the advantages of neutral atoms in a modified unconventional Rydberg pumping mechanism, combined with controlled dissipation, we achieve a three-dimensional GHZ state with a fidelity surpassing 99\% through multiple pump and dissipation cycles. This innovative approach paves the way for experimentally feasible, deterministic preparation of high-dimensional GHZ states in Rydberg atom systems, thereby advancing the capabilities of quantum information processing.
\end{abstract}

\maketitle
Entanglement within higher-dimensional quantum systems introduces more intricate and complex correlations among the states of the entangled particles.\cite{PhysRevLett.89.240401,PhysRevLett.94.220501,PhysRevLett.110.030501,10.1038/nature22986,doi:10.1126/science.aar7053,PhysRevApplied.11.064058,PhysRevLett.123.070505,PhysRevLett.125.230501,erhard2020advances} 
They can prove beneficial for quantum computation and simulation,\cite{lanyon2009simplifying,doi:10.1126/science.1173440,kaltenbaek2010optical,PhysRevA.96.012306} enhance security in quantum key distribution protocols,\cite{PhysRevLett.88.127902,PhysRevA.69.032313,PhysRevA.88.032309} provide increased capacity and noise resistance in quantum communications,\cite{PhysRevA.61.062308,PhysRevA.71.044305,PhysRevLett.96.090501,PhysRevLett.98.060503,PhysRevLett.111.010501,doi:10.1126/sciadv.1701491} and produce stronger violations of Bell-type inequalities. \cite{PhysRevLett.85.4418,PhysRevA.64.024101} These remarkable advantages of high-dimensional quantum entanglement underscore its significance as a valuable resource for quantum information processing, which has led researchers to actively explore methods for preparing high-dimensional entangled states in various physical systems.\cite{PhysRevA.89.012319,PhysRevA.89.052313,malik2016multi,Hebbache_2016,PhysRevA.96.053849,doi:10.1126/science.aar7053,Guo:18,10.1038/s41567-018-0347-x,10.1038/s41534-020-00318-6,10.1038/s41567-022-01658-0,10.1038/s41467-023-37375-2}

The high-dimensional GHZ state is defined by the expression $|\textup{GHZ}(N,d)\rangle=\sum_{k=0}^{d-1}|k\rangle^{\bigotimes N}/\sqrt{d}$, involving $N$ particles, with each particle encoding a qudit of dimension $d$.
This state is of notable significance in various applications, such as quantum key distribution\cite{PhysRevA.97.032312,10.1007/s10773-019-04349-4} and programmable qudit-based quantum processors.\cite{10.1038/s41467-022-28767-x}
In the realm of linear optics, only a limited number of theoretical and experimental strategies have been proposed to create high-dimensional GHZ states. These strategies take advantage of various degrees of freedom of photons as primary carriers of information.\cite{PhysRevLett.118.080401,PhysRevLett.119.240403,erhard2018experimental,doi:10.1073/pnas.1815884116,PhysRevLett.126.230504,Bell_2022,nature.very.large,Xing:23}  One such scheme suggests superposition of photon pairs with different crystals and alignment of photon paths to engineer multiphoton GHZ states in higher dimensions \cite{PhysRevLett.118.080401}  with an efficiency of $E=d/[(N\cdot d)/2]^{N/2}$.  Subsequently, the same group, drawing inspiration from the computer algorithm MELVIN,\cite{PhysRevLett.116.090405} develops a novel multiport design to generate a genuine three-dimensional GHZ state. This approach requires two pairs of three-dimensionally entangled photon pairs as the entanglement resource.\cite{erhard2018experimental} 
Another theoretical approach is proposed, employing a Fourier transform matrix within a linear optical circuit in a photonic system to generate high-dimensional states.\cite{PhysRevLett.126.230504} While this successfully addresses the scalability issue present in postselected schemes,\cite{PhysRevLett.118.080401,erhard2018experimental} the success probability diminishes exponentially with an increase in dimension and photon number. For example, the success probability for generating the $|\textup{GHZ}(3,3)\rangle$ state is merely $10^{-10}$. Even with optimization efforts, it can only be enhanced to $10^{-4}$.

\begin{figure*}
\centering  
\subfigure{(a)}{
\label{fig:subfigc}
\includegraphics[width=0.42\linewidth]{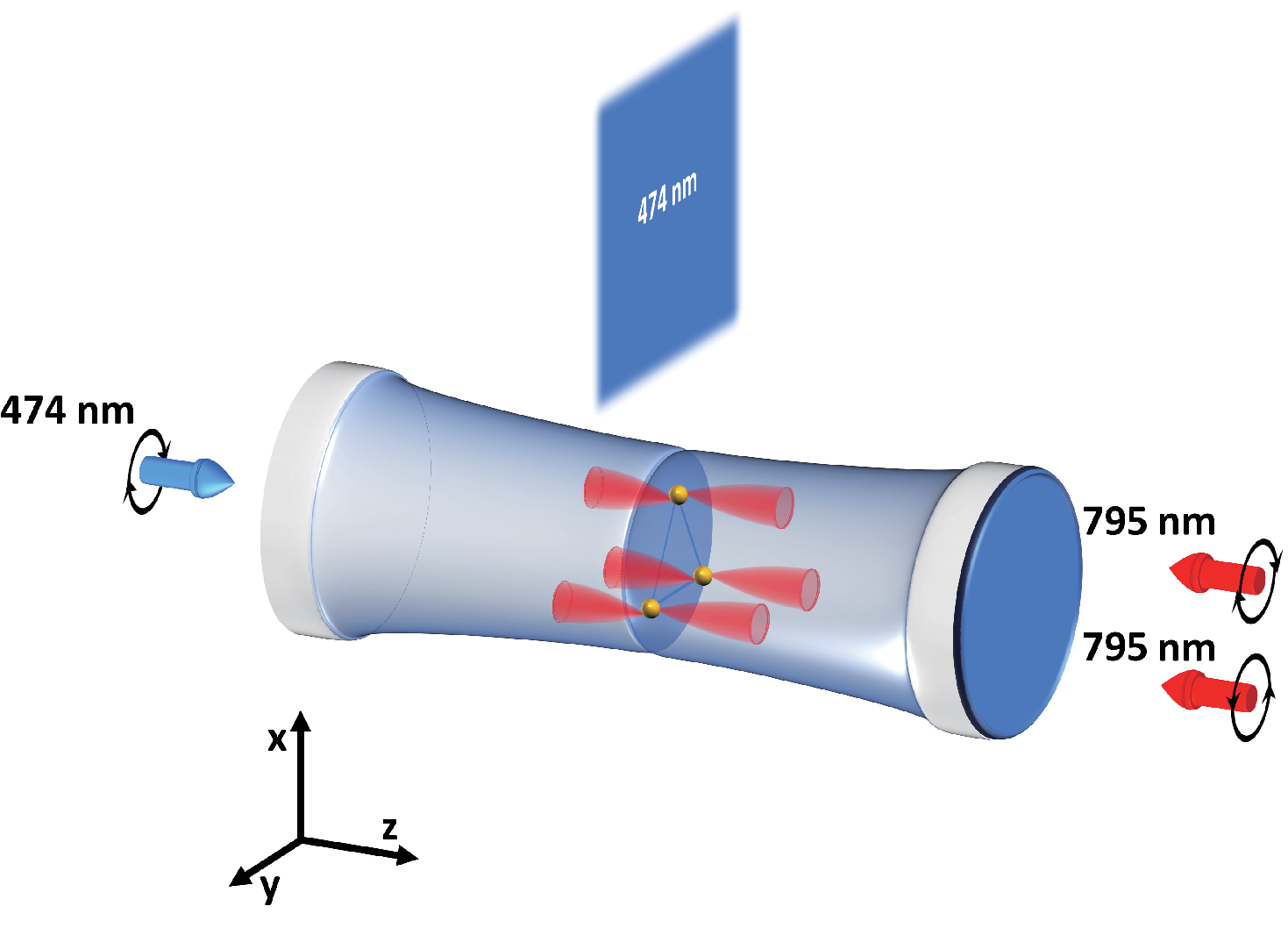}}
\subfigure{(b)}{
\label{fig:subfigd}
\includegraphics[width=0.5\linewidth]{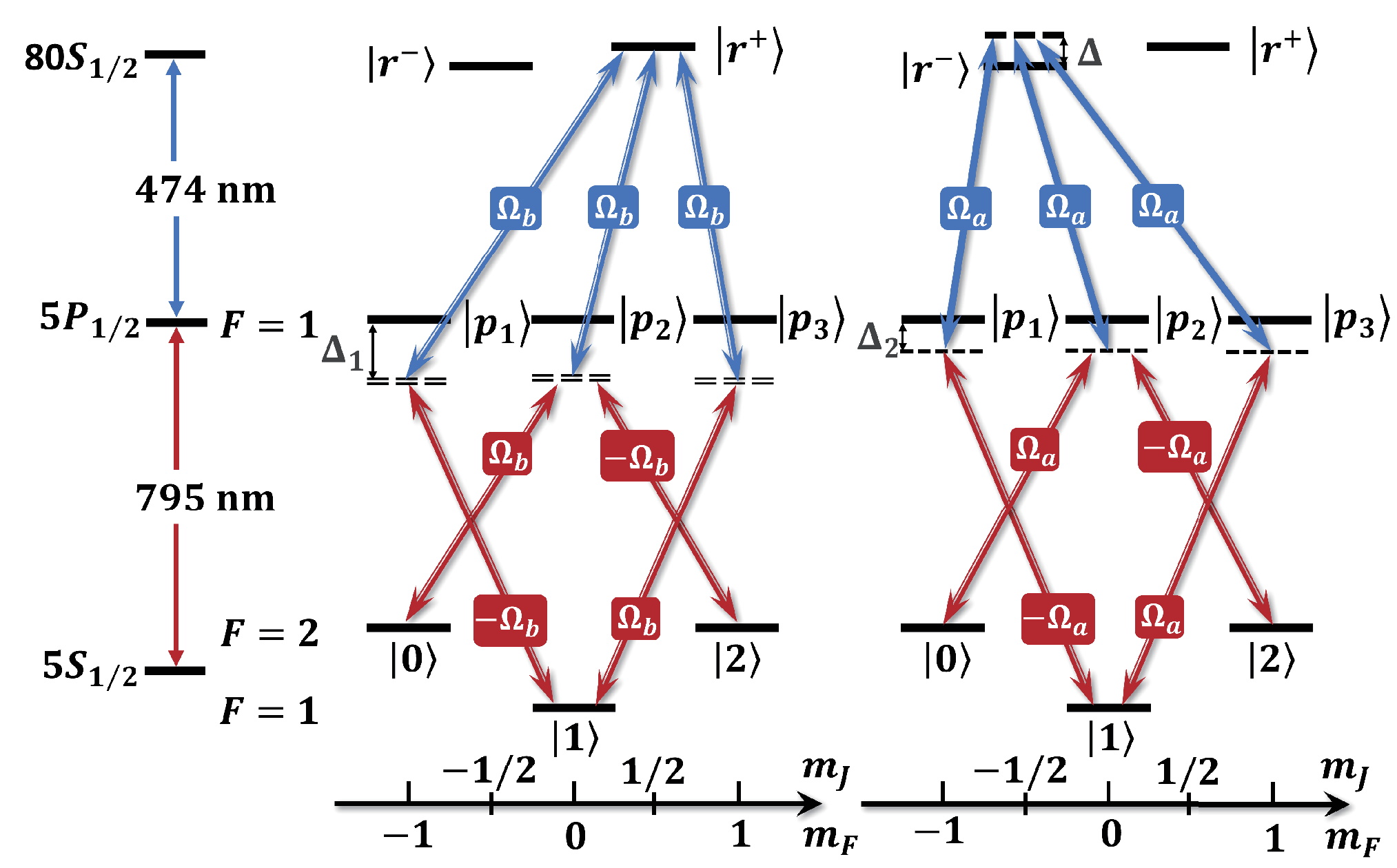}}
\caption{\label{level}Schematic for a potential experimental setup. (a) The atoms are confined using optical tweezers and arranged in an equilateral triangle within the $x$–$y$ plane. 
The use of counterpropagating circularly polarized light beams with various wavelengths significantly reduces the influence of the Doppler effect during atomic excitation from the ground state to the Rydberg state. Simultaneously, co-propagating circularly polarized light beams with the same wavelength prevents the creation of unpredictable relative phases between ground states.
 (b) Depiction of pertinent levels in $^{87}$Rb and the laser-atom interaction model illustrate the implementation of the modified unconventional Rydberg pumping.}
\end{figure*}

The characteristics of the linear optical system determine that all the schemes based on it are nondeterministic. Therefore, to prepare a deterministic high-dimensional GHZ state suitable for quantum information processing, it is necessary to leverage the coupling of light and matter\cite{Bell_2022} or explore the development of alternative matter qubits.
An experimental demonstration and certification of a three-dimensional GHZ state with superconducting transmon qutrits was recently reported.\cite {PhysRevApplied.17.024062} This scheme is also grounded in the quantum circuit employed in MELVIN, \cite{PhysRevLett.116.090405} and the experiment is executed using IBMQ quantum processors.
Nevertheless, due to the inevitable interaction between the superconducting quantum processor and its surroundings, various decoherence channels emerge, particularly qubit crosstalk, a common occurrence in superconducting transmon devices. This results in a fidelity of approximately 76\% for the final prepared quantum entangled state.

In this Letter, our objective is to deterministically prepare a high-dimensional GHZ state by integrating quantum reservoir engineering into a neutral-atom platform which provides a controlled manner for trapping, cooling, and manipulating atoms. \cite{PhysRevLett.114.173002,10.1063/5.0069195,10.1063/5.0086357,10.1063/5.0137127}
This approach converts the decoherence factors, such as atomic spontaneous emission, of the quantum system into valuable resources and operates independently of the preparation of the initial state, which represents a fundamentally distinct scenario from previous preparations of high-dimensional GHZ states in linear optical systems and superconducting systems.

The neutral atom exhibits a plethora of stable hyperfine ground states, with examples including 8 for Rubidium and 16 for Cesium. Leveraging these states proves effective in expanding the dimensionality of a qudit.\cite{RevModPhys.82.2313} By engineering pairwise additive interactions between excited Rydberg states, entangling operations can be facilitated, potentially simplifying the scaling of quantum systems.\cite{PhysRevLett.85.2208,PhysRevLett.87.037901,PhysRevLett.93.063001,PhysRevLett.98.023002,urban2009observation,gaetan2009observation,PhysRevLett.104.013001,PhysRevLett.104.223002,PhysRevA.82.033412,PhysRevLett.105.160404,doi:10.1126/science.aax9743,Browaeys2020,doi:10.1126/science.abg2530,Scholl2021,Graham2022,bluvstein2022quantum,Srakaew2023,Evered2023}  By incorporating the inherent advantages of neutral atoms into the modified unconventional Rydberg pumping, as outlined in the supplementary material, we can effectively stabilize the system within a high-dimensional GHZ state throughout multiple pump and dissipation cycles. To elucidate the mechanism of the current scheme more comprehensively, this work predominantly focuses on a detailed discussion of the preparation of three-dimensional GHZ states. Concerning GHZ states with higher dimensions, their extension can be readily achieved, in principle.

The requisite experimental configuration for our proposed scheme is illustrated in Fig.~\ref{level}. We envision three $^{87}$Rb atoms confined using optical tweezers and arranged in an equilateral triangle within the $x$–$y$ plane, where the interatomic distance ensures effective operation of the van der Waals (vdW) regime. Each qutrit is encoded into three hyperfine ground states: $|0\rangle=|5S_{1/2}, F=2, m_F=-1\rangle$, $|1\rangle=|5S_{1/2}, F=1, m_F=0\rangle$, and $|2\rangle=|5S_{1/2}, F=2, m_F=1\rangle$. Simultaneously, we introduce the degenerate subspace $\{|r^{\pm}\rangle=|80S_{1/2}, m_J=\pm1/2\rangle\}$ of the highly excited Rydberg state to modulate the interaction between the atoms.
The incorporation of an additional $s$-orbital Rydberg sublevel alters the original Hamiltonian of the system, rendering it time-independent.\cite{PhysRevA.98.062338} This adjustment facilitates the implementation of rigorous and efficient numerical simulations. In the two-photon process mediated by the $5P_{1/2}$ states, the Rabi frequencies and the red (blue) detuning for the transition from the ground states $|0\rangle$, $|1\rangle$, and $|2\rangle$ to $|r^+\rangle$ are indicated by $\pm\Omega_b$ and $\Delta_1$, respectively. Similarly, the Rabi frequencies and red (blue) detuning for the transition from ground states to $|r^-\rangle$ are denoted as $\pm\Omega_a$ and $\Delta_2$ ($\Delta_2+\Delta$). Furthermore, coupling between the $5P_{1/2}$ state and the two Rydberg states is achieved through distinct polarized light sources, thus minimizing the influence of light with varying frequencies and intensities. For simplification purposes, we temporarily overlook the influence of atomic position on the relative phase between individual atoms, reserving this discussion for the supplementary material.

\begin{figure}
\centering
\includegraphics[width=1.0\linewidth]{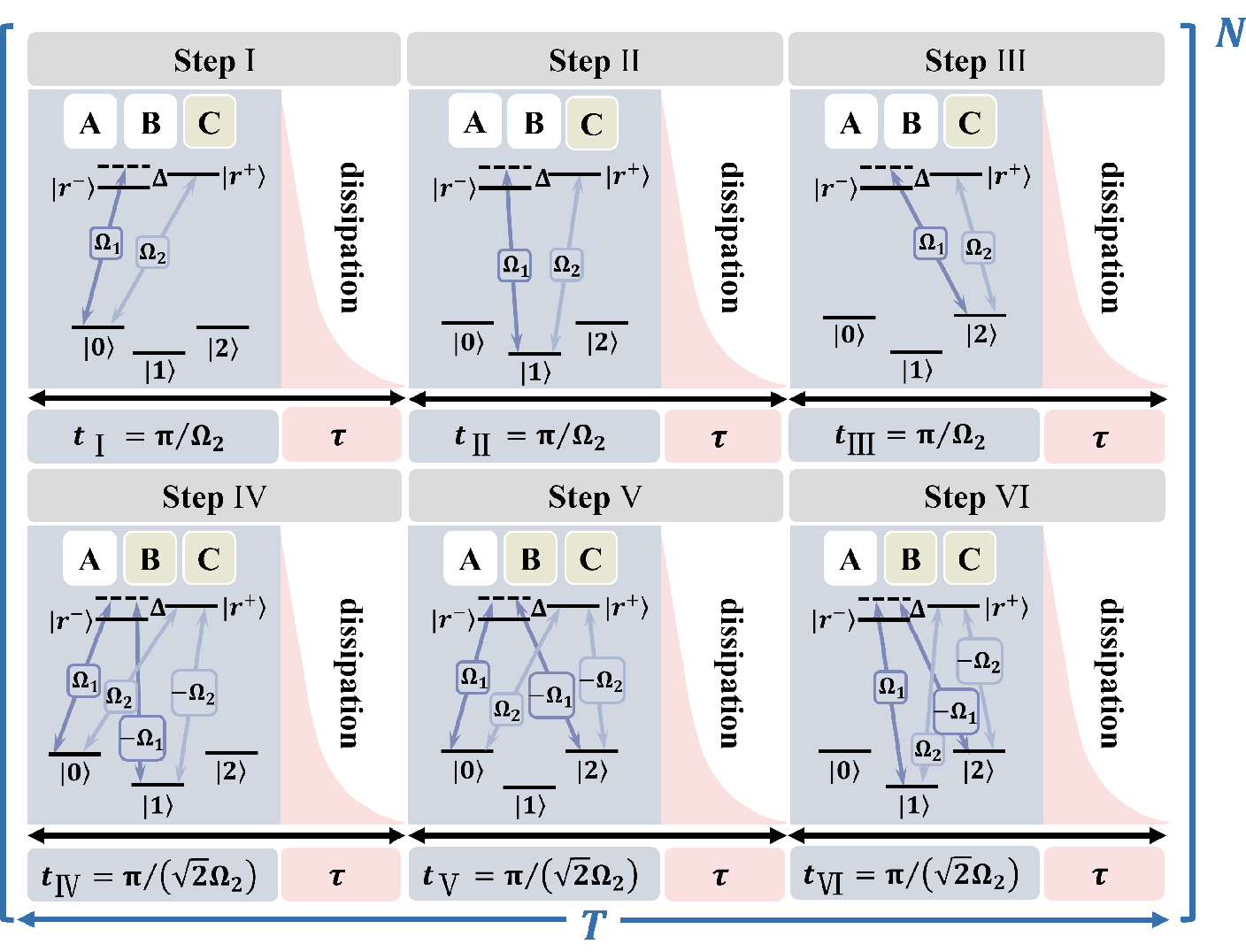}
\caption{\label{LCT6} The flowchart illustrates the step-by-step process for the preparation of the target state $|\textup{GHZ(3,3)}\rangle$. The light green shading in \textbf{B} or \textbf{C} indicates the excitation of some of these states. In the middle section, lasers are applied to different ground states at each step. The lower part of this section details the timing of the driving in each step, denoted as $\pi/\Omega_2$ or $\pi/(\sqrt{2}\Omega_2)$. The light pink segment portrays the dissipation occurring in each step, and its lower portion specifies the dissipation time $\tau$.}
\end{figure}

Aligned with the laser cooling approach, our methodology comprises repetitive cycles of pumping and dissipation, as shown in Fig.~\ref{LCT6}. Within each cycle, we systematically perform six distinct coherent operations, each succeeded by a consistent dissipative operation. Significantly, at each stage, the utilized laser serves the purpose of collectively exciting three atoms, obviating the need for individual addressing. 

In the interaction picture, the corresponding Hamiltonian of the coherent operation at step $i$ can be described by ($\hbar$=1)
\begin{eqnarray}
{\hat H}^{(i)}&=&\sum_{j=1}^{3}\frac{\alpha^i_j}{2}(\Omega_{1}|r^-_{j}\rangle\langle a_{j}^{i}|+\Omega_{2}|r^+_{j}\rangle\langle a_{j}^{i}|)+{\rm H.c.}-\Delta|r^-_{j}\rangle\langle r^-_{j}|\nonumber\\&&-\sum_{k\neq j}[\frac{C_6}{2R^6}(|r^+_{j}\rangle\langle r^+_{j}|\otimes|r^+_{k}\rangle\langle r^+_{k}|+|r^-_{j}\rangle\langle r^-_{j}|\otimes|r^-_{k}\rangle\langle r^-_{k}|)\nonumber\\&&+\frac{C’_6}{R^6}|r^+_{j}\rangle\langle r^+_{j}|\otimes|r^-_{k}\rangle\langle r^-_{k}|],
\label{eq:1}
\end{eqnarray}
where we have obtained the effective single-photon transition with Rabi frequencies $\Omega_1={\Omega_a^2 (2\Delta_2+\Delta)}/[{4 \Delta_2 (\Delta_2+\Delta)}]$ and $\Omega_2=\Omega_b^2 /({2 \Delta_1})$ between ground states and Rydberg states after adiabatically eliminating the intermediate state $|5P_{1/2}\rangle$ in Fig.~\ref{level}(b). The state $|a_{j}^{i}\rangle$ represents the laser-driven normalized ground state of the $j$th atom, with $\alpha_j^i$ serving as the associated collective enhancement factor. It should be noted that in the first three steps, the parameter $\alpha_j^i$ is set to 1, while for subsequent steps it is adjusted to $\sqrt{2}$. Additionally, the dispersion coefficients $C_6$ and $C'_6$ for the high-lying $s$ state of Rydberg atoms demonstrate isotropy with respect to the polar angle, representing the orientation of the interatomic connection to the quantization axis. Specifically, for $n=80$, their values are $C_6/(2\pi)=-4160$ GHz $(\mu\mathrm{m})^6$ and $C'_6/(2\pi)=-4213$ GHz $(\mu\mathrm{m})^6$. Consequently, the interaction strength between Rydberg atoms can be tuned by altering the atomic spacing $R$. In the scenario of $C'_6/R^6=-\Delta$, and under the limit of $\Delta\gg\Omega_{1}\gg\Omega_2$, an effective Hamiltonian can be derived as (see the supplementary material for details)
\begin{eqnarray}
\hat{H}_{\textup{eff}}^{(i)}=&&\frac{\alpha^i\Omega_{2}}{2}\sum_{m, n}(|r^+_1\rangle\langle a_{1}^{i}|P_2^mP_3^n+P_1^m|r^+_2\rangle\langle a_{2}^{i}|P_3^n\nonumber\\&&+P_1^mP_2^n|r^+_3\rangle\langle a_{3}^{i}|)+\mathrm{H.c.},
\label{eq7}
\end{eqnarray}
where $P^m$ and $P^n$ represent the projection operators onto two distinct bases that are orthogonal to the state $|a^{i}\rangle$ within the atomic ground state. 
Following Eq.~(\ref{eq7}), the system allows only a single excitation process associated with the ground state, allowing the ground state $|a^i\rangle$ to undergo excitation to the Rydberg state $|r^+\rangle$ at intervals of $\pi/\Omega_2$ during the initial three steps, and with a period of $\pi/(\sqrt{2}\Omega_2)$ during the subsequent steps. The jump operators that characterize each atomic spontaneous emission, taking into account the decay rate, are defined as follows:
\begin{equation}
\hat{L}^{l}=\sqrt{\frac{\Gamma_1}{3}}|l\rangle\langle r^{+}|,\  \  (l=0,1,2).
\end{equation}
Here, the effective decay rate $\Gamma_1$ can be significantly increased beyond the natural linewidth of $|80S_{1/2}\rangle$ for $^{87}$Rb atoms using controlled dissipation techniques. \cite{PhysRevA.96.043411,PhysRevApplied.20.014014,begoc2023controlled} This enhancement facilitates the acceleration of the atomic relaxation process to the desired ground state.

\begin{figure}
\centering
\includegraphics[width=0.97\linewidth]{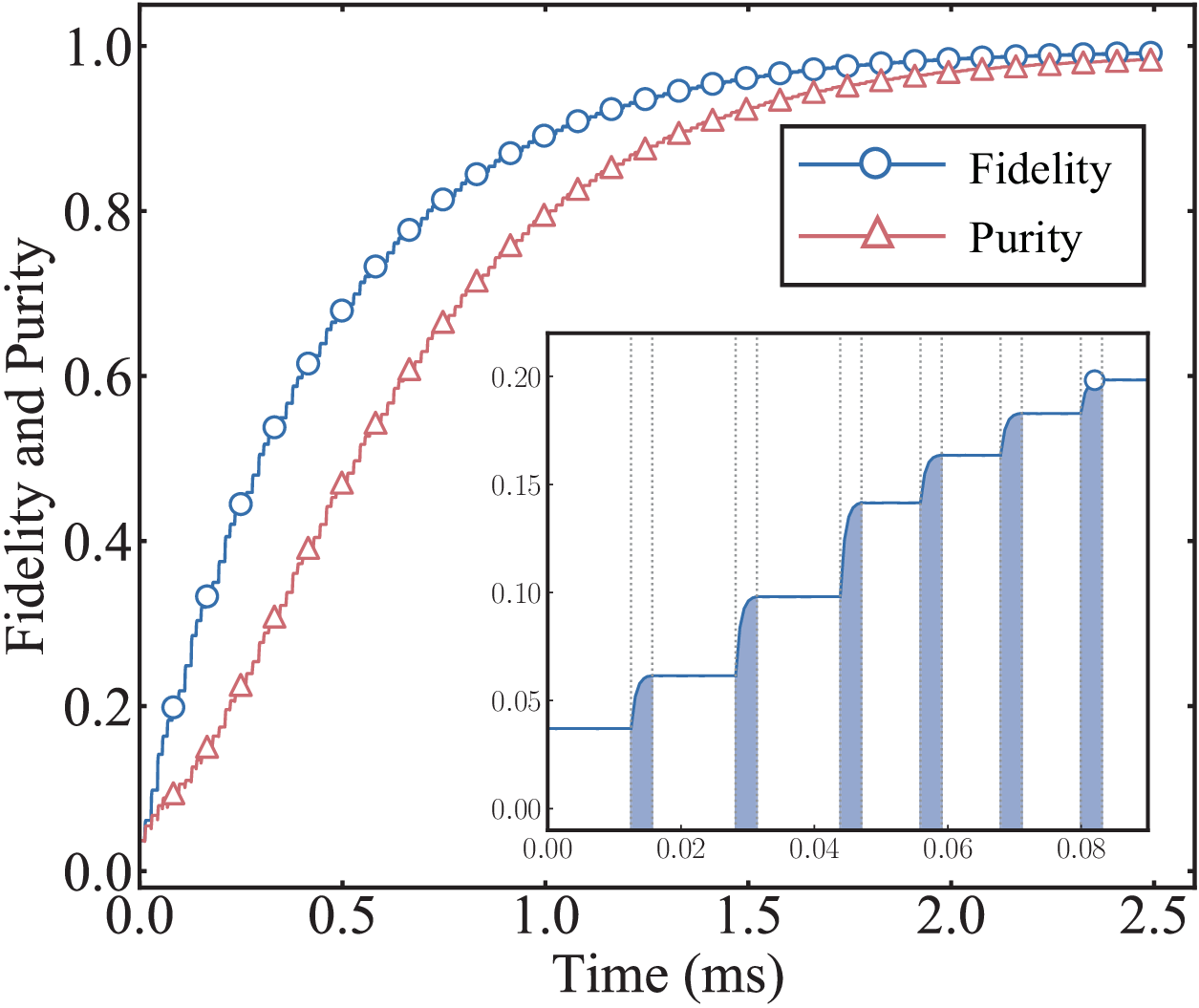}
\caption{\label{Fidelity} The temporal evolutions of fidelity and purity of the target state $|{\textup{GHZ(3,3)}}\rangle$ are illustrated against time (number of cycles) starting from an arbitrary initial state. The selected parameters are $\Omega_1=2\pi\times4$ MHz, $\Omega_2=2\pi\times0.04$ MHz, $\Delta=2\pi\times200$ MHz, $\Gamma_1=2\pi\times0.23$ MHz, and $\tau=3.19~\mu$s. The time required for 30 cycles is approximately 2.49 ms. The inset provides a detailed view of the fidelity in the first circle for each step.}
\end{figure}

To elaborate on the details of the scheme, we partition the ground state space of three atoms into three distinct subspaces. Subspace \textbf{A} is represented by $\{(|000\rangle+|111\rangle+|222\rangle)/\sqrt{3}\}$, which corresponds to the target state $|{\rm GHZ(3,3)}\rangle$. The subspace \textbf{B} $=\{(2|222\rangle-|000\rangle-|111\rangle)/\sqrt{6}, (|000\rangle-|111\rangle)/\sqrt{2}\}$ consists of states that are orthogonal to the target state. Together, subspace \textbf{A} and subspace \textbf{B} form a complete set of subspace $\{|000\rangle, |111\rangle, |222\rangle\}$. Lastly, subspace \textbf{C} encompasses the remaining 24 orthogonal basis states. 
Under the action of Eq.~(\ref{eq7}), the initial three steps illustrated in Fig.~\ref{LCT6} allow one of the three atoms
in the ground state 
to be excited to the Rydberg state $|r^+\rangle$ if and only if the pumped atom is initialized in the $|a^i\rangle$. 
After controlled dissipation, the distribution of each component of the atomic ground state is adjusted. We find that under the assumption of considering the first three steps in general, the stability of subspace \textbf{C} is destroyed and the system will be stabilized to a mixed state composed of states in subspaces \textbf{A} and \textbf{B}.
The operational steps in \uppercase\expandafter{\romannumeral+4}, \uppercase\expandafter{\romannumeral+5}, and \uppercase\expandafter{\romannumeral+6} are analogous to the initial three steps. The impact of these lasers becomes more apparent when we express the bare ground states of all atoms in terms of the dressed states, as deduced in the supplementary material. Let us consider step \uppercase\expandafter{\romannumeral+4} as an example.
The state $|a^{IV}\rangle=(|0\rangle-|1\rangle)/\sqrt{2}$ can be pumped to $|r^+\rangle$ after a period $\pi/(\sqrt{2}\Omega_2)$, and then  $|r^+\rangle$ decays towards $|0\rangle$, $|1\rangle$, and $|2\rangle$ independently.
This process decreases the population of $|a^{IV}\rangle$ and improves the population of $|a^{IV}_\perp\rangle=(|0\rangle+|1\rangle)/\sqrt{2}$. Consequently, within the subspaces of \textbf{A} and \textbf{B}, all mixed states with components like $(|000\rangle-|111\rangle)/\sqrt{2}=(|a^{IV}a^{IV}a^{IV}\rangle+|a_\perp^{IV}a_\perp^{IV}a^{IV}\rangle+|a_\perp^{IV}a^{IV}a_\perp^{IV}\rangle+|a^{IV}a_\perp^{IV}a_\perp^{IV}\rangle)/2$ are selectively pumped out. In contrast, components containing $(|000\rangle+|111\rangle)/\sqrt{2}=(|a_\perp^{IV}a_\perp^{IV}a_\perp^{IV}\rangle+|a^{IV}a^{IV}a_\perp^{IV}\rangle+|a^{IV}a_\perp^{IV}a^{IV}\rangle+|a_\perp^{IV}a^{IV}a^{IV}\rangle)/2$ in the mixed states persistently increase.

Similarly, in steps \uppercase\expandafter{\romannumeral+5} and \uppercase\expandafter{\romannumeral+6}, the components of $(|000\rangle+|222\rangle)/\sqrt{2}$ and $(|111\rangle+|222\rangle)/\sqrt{2}$ also experience an increase. Hence, it becomes evident that after an evolutionary period, only the population of the target state $|{\rm GHZ(3,3)}\rangle$ remains unaffected by laser pumping, gradually increasing through controlled spontaneous emission. This is illustrated by the light blue shading in the inset of Fig.~\ref{Fidelity}. After multiple cycles, the initial population in any state will be transferred to the target state. To underscore the robustness of the scheme against variations in the initial state, we intentionally choose a completely mixed state, 
$\rho(0)=\sum_{i,j,k=0,1,2}|ijk\rangle\langle ijk|/27$,
as the initial condition. The fidelity, defined by the population ${\mathcal{F}}=\langle{\rm GHZ(3,3)}|\rho(t)|{\rm GHZ(3,3)}\rangle$, of the target state
$|{\rm GHZ(3,3)}\rangle$, and the purity ${\mathcal{P}}={\rm Tr}[\rho(t)^2]$ are then plotted against time (number of cycles) in Fig.~\ref{Fidelity}, providing insight into the performance of the scheme. The corresponding parameters, $\Omega_1=2\pi\times4$ MHz, $\Delta=50\Omega_{1}$ and $\Omega_2=0.01\Omega_1$, are chosen to satisfy the conditions of unconventional Rydberg pumping. After 30 cycles, the fidelity of $|{\textup{GHZ(3,3)}}\rangle$ reaches 99.1\%, and the entire process is completed in approximately 2.49~ms, showcasing the efficacy of the chosen parameters within the specified temporal constraints.

In the context of Rydberg atom platform experiments, various imperfections are inherent, encompassing factors such as the random thermal motion of atoms within the optical tweezers, spontaneous emission from the intermediate state, laser phase noise, and the Doppler effect.\cite{PhysRevA.72.022347,PhysRevA.97.053803,PhysRevA.99.043404} These imperfections significantly impact the coherence between the ground state and the excited state of Rydberg atoms. Addressing these factors is essential for overcoming challenges and realizing efficient quantum information processing in the Rydberg atomic system. In our proposed scheme, the spontaneous emission of atoms has been transformed into a valuable resource, and the remaining factors can be attributed to the influence of finite temperature. Specifically, the dephasing effect induced by laser phase noise can be equivalently regarded as the Doppler effect.

Given the atomic arrangement illustrated in Fig.\ref{level}(a), for atoms at a temperature $T_a$, the time-averaged variances of position and momentum are approximated as $\langle x^2 \rangle = \langle y^2 \rangle = k_B T_a / (\omega^2 m)$ and $\langle v_x^2 \rangle = \langle v_y^2 \rangle = k_B T_a / m$. Here, $k_B$ represents the Boltzmann constant, $\omega=\omega_{x(y)}$ is the oscillation frequency of the trap, and $m$ is the atomic mass. Longitudinal position fluctuations combine in quadrature, making a lesser contribution to distance fluctuations.\cite{10.1038/nature24622} Therefore, the deviation of the interatomic distance is expressed as $\sigma = \sqrt{2k_B T_a / (\omega^2 m)}$. For a typical trap frequency of $\omega=2\pi\times90~$kHz,\cite{PhysRevLett.110.263201} an atomic temperature of $10~\mu$K results in $\sigma=77.37~$nm. The lower atomic temperatures experimentally obtained, $T_a=5.2~\mu$K and $T_a=1~\mu$K, lead to reduced position fluctuations of $\sigma=55.79$~nm and $\sigma=24.47~$nm, respectively.\cite{PhysRevA.105.042430,zhao2023floquet} In a recent experiment, a higher trap frequency with $(\omega_x,\omega_y)=2\pi\times(147,117)~$kHz \cite{chew2022ultrafast} can be used to further suppress fluctuation of the atomic spacing. 

\begin{figure}
\centering
\includegraphics[width=1.0\linewidth]{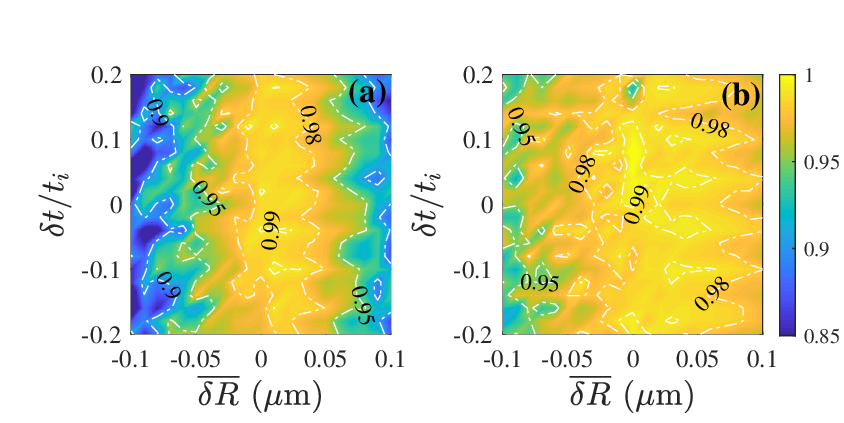}
\caption{\label{erroe22} The influence of distance and timing errors is showcased in two distinct scenarios. In (a), a rectangular pulse with a Rabi frequency of $\Omega_2=2\pi\times0.04$ MHz is considered, whereas in (b), the analysis entails a rectangular pulse with a reduced Rabi frequency of $\Omega_2=2\pi\times0.02$ MHz, achieving a more robust effect.}
\end{figure}

To assess the resilience of our approach to variations in atomic spacing $\overline{\delta R}$ around the target value of $R_0=5.255~\mu$m and to account for timing errors $\delta t$ during coherent operations, we illustrate a contour plot of the target state population after 30 cycles as a function of $\overline{\delta R}$ and $\delta t$ in Fig.~\ref{erroe22}(a). The target state population remains consistently above 90\% within the $\overline{\delta R}$ range of $[-50,50]~$nm. The impact of timing errors ${\delta t}/t_i\in[-0.2,0.2]$ with $t_i$ corresponding to the timing for each step in Fig.~\ref{LCT6}, denoted as $\pi/\Omega_2$ or $\pi/(\sqrt{2}\Omega_2)$ in the population is nonmonotonic, manifesting itself as oscillations.
In the supplementary material, we present evidence demonstrating that selecting a smaller Rabi frequency for the (effective single-photon) resonant laser enhances the resilience of the target state population against variations in atomic spacing, as illustrated in Fig.~\ref{erroe22}(b). While this improvement is attained by extending the duration of the coherent operations and may be accompanied by potential influences of various unknown decoherence factors, it nonetheless stands as a viable option to effectively tackle challenges arising from fluctuations in atomic spacing.

\begin{figure}
\centering\includegraphics[width=0.97\linewidth]{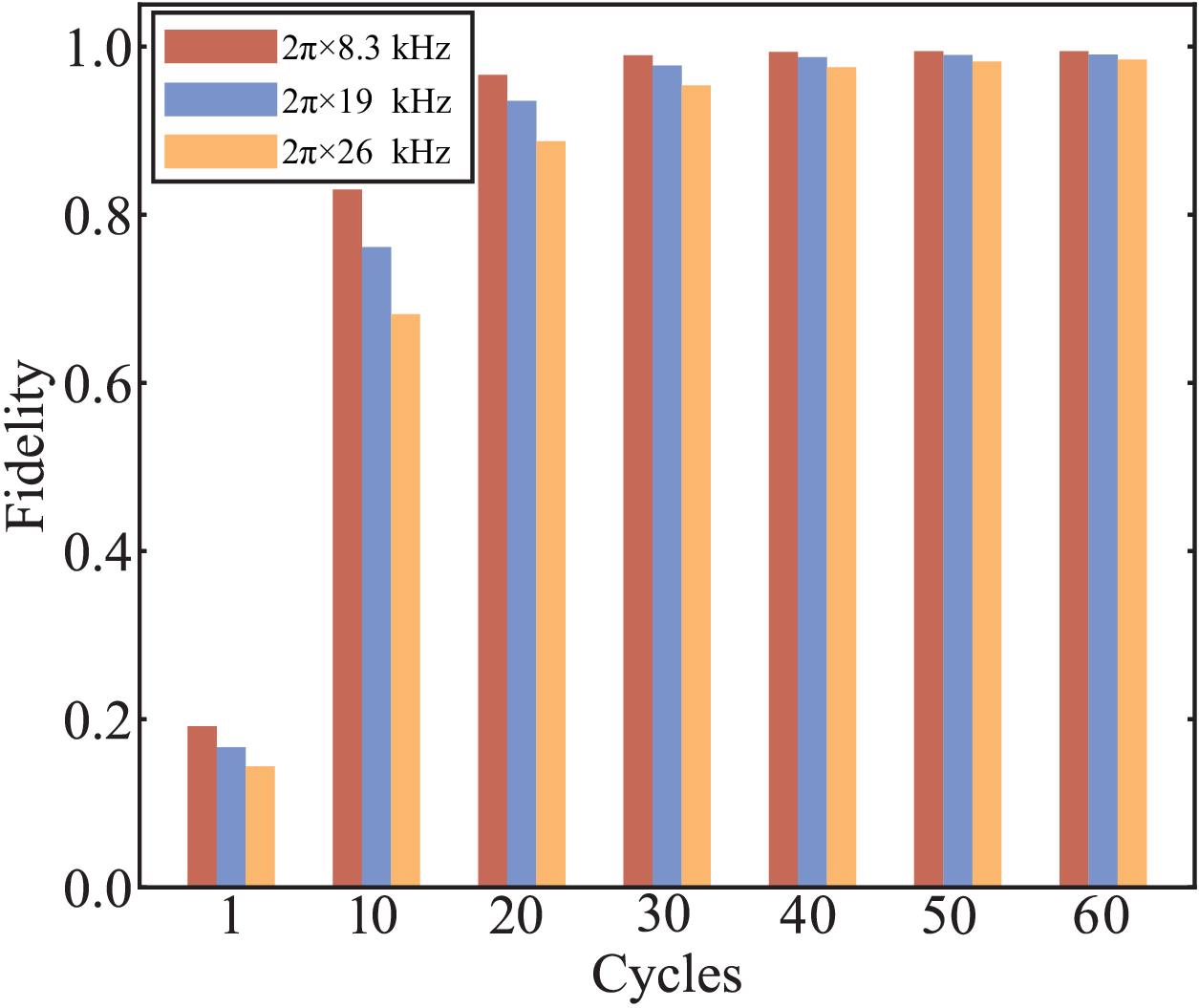}
\caption{\label{Doppler} The influence of Doppler effect. The driving field with Rabi frequency $\Omega_2$ have an extra detuning of $2\pi\times(8.3,19,26)$ kHz.}
\end{figure}

In our setup, where linearly polarized light and circularly polarized light are employed, the effective wave vector for the two-photon excitation from the ground state to the excited state of the Rydberg atom is conservatively estimated to be $k_{\rm eff}\approx1.54\times10^7~{\rm m^{-1}}$ at maximum (orthogonal beams). 
As a result, the atom is subjected to an extra detuning of the excitation laser, which is represented as a random variable with a Gaussian probability distribution centered around zero and a standard deviation of $k_{\rm eff}\sqrt{k_BT_a/m}\approx 2\pi\times(76, 55, 24)~$kHz. This effect, however, has no effect on the two-photon off-resonance pathway from the ground state to the Rydberg state $|r^-\rangle$. The higher Rabi frequency $\Omega_1$ associated with this process is attributed to this,\cite{PhysRevA.97.053803} and the frequency detuning induced by the Doppler effect is negligible compared to the detuning of $\Delta$ in our scheme.
On the other hand, for the resonant Rydberg atomic transition, the counterpropagating 474-nm and 795-nm wavelengths yield a small effective wave vector of $k_{\rm eff}\approx5.35\times10^6~{\rm m^{-1}}$. This implies a corresponding $k_{\rm eff}\sqrt{k_BT_a/m}\approx 2\pi\times(26, 19, 8.3)~$kHz.
Applying this factor to the resonant Rydberg atomic transition using standard random sampling and average processing is unquestionably time-consuming. Within this particular framework, we simplify the procedure by replacing the detuning frequency with the standard deviation previously mentioned and qualitatively determine the influence of the Doppler effect on our scheme. In Fig.~\ref{Doppler}, we track the progression of target state fidelity as influenced by these three parameters over multiple cycles, respectively. We permit some populations to remain in the ground state under each coherent control due to the periodic pumping and dissipation of the scheme itself, these residual populations can be progressively transferred to the target entangled state of the atom in the subsequent cyclic operation. Consequently, we observe that it is still possible to attain a greater population of the target state by increasing the number of cycles. 

In conclusion, we have presented a scheme for the deterministic preparation of high-dimensional GHZ states, leveraging periodic pumping and dissipation in a neutral atom system. This method remains independent of the initial state preparation and exhibits resilience to fluctuations in atomic spacing and timing errors. Moreover, it proves advantageous in mitigating the impact of the Doppler effect, as a higher fidelity target state can still be achieved by increasing the number of cycles.
While extending our approach to a high-dimensional GHZ state is straightforward, it necessitates additional coherent manipulations and evolutionary cycles. However, the spatial arrangement structure limitations between Rydberg atoms pose a challenge, rendering the current experimental parameters insufficient to guarantee the robust creation of a high-dimensional GHZ state with more than three particles. We anticipate that ongoing developments in atomic manipulation technology and advancements in atomic cooling technology will pave the way for the realistic preparation of a high-dimensional multi-particle GHZ state in the future.

See the supplementary material for a detailed description of the effective two-level system, the effective Hamiltonian, the controlled spontaneous emission of Rydberg states, the relative phases of atoms induced by wave vectors, and the extension of three-dimensional GHZ states into higher dimensions.

This work is supported by the National Natural Science
Foundation of China (NSFC) under Grant No. 12174048. W.L. acknowledges support from the EPSRC through Grant No.
EP/W015641/1, and the Going Global Partnerships Programme of the British Council (Contract No. IND/CONT/G/22-23/26).

\vspace{12pt}
\section*{AUTHOR DECLARATIONS}
\subsection*{Conflict of Interest}
The authors have no conflicts to disclose.

\vspace{12pt}
\subsection*{Author Contributions}
{\bf Yue Zhao}: Numerical simulation (equal); 
 Writing-original draft (equal). {\bf Yu-Qing Yang}: Numerical simulation (equal); 
 Writing-original draft (equal). {\bf Weibin Li}: Writing-review \&
editing (equal); Supervision (equal). {\bf Xiao-Qiang Shao}: Conceptualization; 
Funding acquisition; Writing-review \&
editing (equal); Supervision (equal). 

\vspace{12pt}
\section*{DATA AVAILABILITY}
The data that support the findings of this study are available from the corresponding authors upon reasonable request.

\vspace{12pt}
\section*{REFERENCES}

\bibliography{ref}
\end{document}